\documentclass[epsf,useAMS,usenatbib,usegraphicx]{mn2e}
\usepackage{epsf}
\usepackage[usenames]{color}

\usepackage[usenames]{color}

\title[Magnetic stars from a FEROS cool Ap star survey ]{Magnetic stars from a 
FEROS cool Ap star survey\thanks{Based on observations collected at the European 
Southern Observatory, La Silla, Chile, as part of programmes 078.D-0080(A), 080.D-
0191(A), 081.D-2002(A), 082.D-0061(A), 083.D-0034(A), 084.D-0067(A) }}

\author[Elkin et al.]{V.G.~Elkin$^{1}$, D.W.~Kurtz$^{1}$, C.~Nitschelm$^{2}$ \\
$^{1}$Jeremiah Horrocks Institute, University of Central Lancashire, Preston 
PR1\,2HE, UK\\
$^{2}$Instituto de Astronom\'ia, Universidad Cat\'olica del Norte, Antofagasta, 
Chile \\}

\begin{document}

\maketitle

\begin{abstract}
New magnetic Ap stars with split Zeeman components are presented. These stars were 
discovered from observations with the FEROS spectrograph at the ESO 2.2-m 
telescope. Fifteen new magnetic stars are analysed here. Several stars with very 
strong magnetic fields were found, including HD\,70702 with a 15-kG, and 
HD\,168767 with a 16.5-kG magnetic field strength measured using split Zeeman 
components of spectral lines and by comparison with synthetic calculations. The 
physical parameters of the stars were estimated from photometric and spectroscopic 
data. Together with previously published results for stars with strong magnetic 
fields, the relationship between magnetic field strength and rotation period is 
discussed.
\end{abstract}

\begin{keywords} stars: magnetic fields - stars: chemically peculiar - techniques: 
spectroscopic

\end{keywords}

\section{Introduction}
\label{sect:intro}

Stars with strong global magnetic fields are found along the main sequence among 
chemically peculiar (CP or Bp/Ap/Fp) stars. They range over spectral type from 
mid-F to early B types up to the terminal age main sequence. Magnetic fields 
produce significant effects in the atmospheres of peculiar stars. The interaction 
of the magnetic field with the plasma in stellar atmospheres is one of the most 
challenging problems in stellar astrophysics. The peculiarities of the CP stars 
are caused by unusual chemical abundances in their atmospheres. The strong 
magnetic field stabilises their outer stellar layers and creates the conditions to 
support atomic diffusion separation of chemical elements (e.g. 
\citealt{Michaud70}; \citealt{Michaud80}). Radiative accelerations and 
gravitational settling in the magnetised atmosphere produce inhomogeneous surface 
distributions and vertical stratification (\citealt{BaMi91}; \citealt{Babel94}).

The magnetic field strength detected in magnetic stars varies from one star to 
another. The strongest magnetic field detected is 34\,kG in HD\,215441 
(\citealt{babcock60}; \citealt{Preston69}), which belongs to the hotter Si type of 
Ap/Bp stars, while for cooler Ap stars the magnetic field reaches 30\,kG 
(\citealt{Freyhammer08}; \citealt{Elkin10b}). Observed magnetic fields in Ap/Bp 
stars vary with rotational period and are described by the oblique rotator model. 
In most magnetic stars to first approximation the magnetic field geometry may be 
considered as a simple dipole, while more precise observations and detailed 
analyses 
can reveal more complex configurations of magnetic field structure.

For a fraction of magnetic stars high resolution spectroscopic observations have 
allowed straightforward measurements of the magnetic field modulus. This 
happens when the star rotates slowly or is observed at low inclination angle, so 
has narrow spectral lines. Fortunately, peculiar stars show lower rotational 
velocities compared with normal stars of the same spectral class. \citet{Mathys97} 
presented comprehensive analyses of stars with magnetic field moduli determined 
from resolved Zeeman components. Since that publication more stars with strong 
magnetic fields and resolved components have been discovered.

In a high resolution spectroscopic survey of cool peculiar stars compiled from a 
list by \citet{martinez93} based on the Michigan Spectral Catalogues, we also have 
found stars with resolved Zeeman components. The results from our first observing 
set with FEROS (Fiber-fed Extended Range Optical Spectrograph) were presented 
in \citet{Freyhammer08}. Further observations of stars from our survey, and 
analyses of their spectra, reveal several more stars with strong magnetic fields. 
One of these, BD\,+0$^\circ$4535, has a strong magnetic field reaching 21\,kG that 
was described by \citet{Elkin10a}. This paper presents fifteen new magnetic Ap 
stars with sharp lines and with Zeeman splitting detected with FEROS observations.

\section{Observations}

The spectra analysed in this paper were obtained with the FEROS echelle 
spectrograph at the La Silla 2.2-m telescope of the European Southern Observatory 
(ESO). The FEROS spectra have a resolution of $R = 48000$ and a wavelength range 
from $\lambda\lambda\,3500 - 9200$\,\AA. Some of the stars were also observed with 
the ESO Very Large Telescope (VLT) at Paranal Observatory using the Ultraviolet 
and Visual Echelle Spectrograph (UVES) installed at Unit Telescope 2 (UT2). UVES 
spectra covered the range $\lambda\lambda\,4970 - 7010$\,\AA, with a small gap 
near 6000\,\AA\ caused by the space between the two CCDs of the spectrograph.

For reduction of spectroscopic observations and extraction of 1D spectra {\small 
ESO-MIDAS} pipelines were used and the 1D spectra were normalised to continuum. 
Some stars were observed two or more times while most were observed only once.

\section{Analysis of the observational data}

\subsection{Magnetic field determination}

Experimental determination of magnetic fields in Ap stars is based on measurement 
of the Zeeman effect in spectral lines. The line of Fe\,\textsc{ii} 6149.258\,\AA\ 
with a convenient doublet Zeeman configuration (see, e.g., \citealt{Mathys97}) was 
used for magnetic determinations. The doublet structure of the Fe\,\textsc{ii} 
6149.258\,\AA\ line allowed us to select and directly measure magnetic field 
strengths greater than about 3\,kG with the FEROS spectra. Only partial Zeeman 
splitting, or just a hint that it may be present in line broadening, is seen for 
smaller field strengths. The peculiarity level and type vary significantly from 
one star to another. The Fe\,\textsc{ii} 6149.258\,\AA\ line is a blend with 
several other lines that can have significant strength for some stars. Other lines 
with large Land\'e factors are thus needed to test for magnetic splitting or 
broadening. Calculations of synthetic spectra using the {\small SYNTHMAG} code by 
\citet{Piskunov99} and comparison with observations are required to distinguish 
the blending effect from magnetic splitting and broadening. A further blending 
problem may occur for very strong fields when the splitting in the Fe\,\textsc{ii} 
6149.258\,\AA\ line is large and the Zeeman components are not clearly visible or 
may be blended.

The distance between shifted Zeeman $\sigma$ components is proportional to the 
value of the mean magnetic field modulus over visible stellar hemisphere  $\langle 
B \rangle$ (e.g. \citealt{Landstreet80}, \citealt{Mathys90}):
\begin{equation}
\Delta\lambda = 9.34 \times 10^{-13} g_{\rm eff}
\langle B \rangle \lambda_{0} ^{2},
\label{eq:B}
\end{equation}
where wavelength is in Angstroms and magnetic field strength is in Gauss. 
Throughout this paper it is this modulus that is meant when we simply refer to the 
magnetic field strength. When the field is strong and the components were 
resolved, each 
component was fitted with a Gaussian. The distance between the central positions 
of the Gaussians was used for the magnetic field determination. Synthetic spectra 
were calculated with the {\small SYNTHMAG} code for different abundances and 
magnetic field strengths to obtain a best fit with the observed spectra. For these 
calculations model atmospheres were obtained from the NEMO database (Vienna New 
Model Grid of Stellar Atmospheres, \citealt{heiteretal02}) and a spectral line 
list 
from the Vienna Atomic Line Database (VALD, \citealt{kupkaetal99}), which includes 
lines of rare earth elements from the DREAM database \citep{biemontetal99}. 
Stellar parameters were estimated for a selection of model atmospheres in the 
ranges of $T_{\rm eff}$ and $\log g$ for these stars. Str\"omgren photometric 
indices \citep{martinez93} and the calibration by \citet{moonetal85} were then 
used for initial estimates of $T_{\rm eff}$ and $\log g$. Synthetic spectra for 
the H$\alpha$ region were then calculated with the {\small SYNTH} code 
\citep{Piskunov92} for different effective temperatures and gravity. Finally, 
synthetic profiles of H$\alpha$ were compared with the observed spectra for a best 
fit and a final determination of $T_{\rm eff}$ and $\log g$. The H$\alpha$ profile 
is not very sensitive to the latter, therefore the $\log g$ estimates were based
 on photometric calibration and are not precise, but are still suitable for our 
purposes.

\subsection{Comments for individual stars}

In this section we present details of further analysis for newly discovered 
magnetic stars. By the standards of stellar magnetic field studies, all of these 
stars are relatively faint and poorly studied. Some limited information about them 
is  present in different stellar catalogues, but most physical parameters are 
unknown. The stars from our list were observed photometrically with the All Sky 
Automatic Survey (ASAS, \citealt{Pojmanski02}). We used this photometry to check 
for rotational variability of each star. For some stars we found a probable 
rotation period using the {\small Period04} package \citep{lenzetal05} and 
discrete Fourier transform of Kurtz (1985).

\subsubsection{HD\,3988}

This is a spectroscopic binary system for which speckle interferometry by 
\citet{White91} does not show any multiplicity. The lower and narrow part of the 
profiles of the Balmer lines, especially H$\alpha$, show binary structure with 
strong and weak components in the double core. The sodium doublet, Na\,\textsc{i} 
5889.951\,\AA\ and 5895.924\,\AA, also shows the spectra of both stars. The lines 
of the two stars show wavelength variability, changing from 1.14\,\AA\ to 
1.29\,\AA\ on different nights of observation. There are insufficient data to 
derive an orbital period. By fitting H$\alpha$ with synthetic spectra we estimate 
the effective temperatures of the primary and secondary components to be 7200\,K 
and 6600\,K, respectively, with an estimated precision of $200 - 300$\,K.

\begin{figure}
\begin{center}
\hfil \epsfxsize 8.2cm\epsfbox{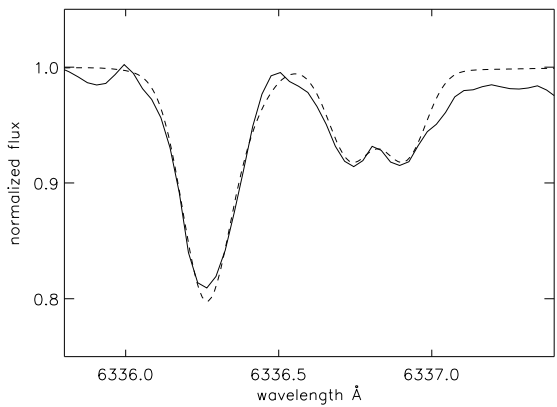}
\caption{\label{sy3988} A spectral region with a magnetically sensitive line for 
HD\,3988. For this and other figures the observed spectrum is shown by a solid 
line, while the synthetic spectrum is presented with a dashed line. The synthetic 
spectrum was calculated for a magnetic field strength of 2.7\,kG. The strongest 
line is Cr\,\textsc{ii} 6336.263\,\AA, while the line with a doublet structure of 
partially split Zeeman components is Fe\,\textsc{i} 6336.823\,\AA.}
\end{center}
\end{figure}

The spectrum is not extremely peculiar and has rather weak lines of rare earth 
elements that belong to one component. The Fe\,\textsc{ii} 6149.258\,\AA\ line 
does not show splitting, but only some broadening that may be connected with a 
magnetic field. Partial splitting is seen in some lines with larger Land\'e 
factors, e.g., the line Fe\,\textsc{i} 6336.823\,\AA. Synthetic calculations give 
good agreement with observations for a magnetic field strength of 2.7\,kG as seen 
in Fig.\,\ref{sy3988}

\subsubsection{HD\,57040}

\begin{figure}
\begin{center}
\hfil \epsfxsize 8.2cm\epsfbox{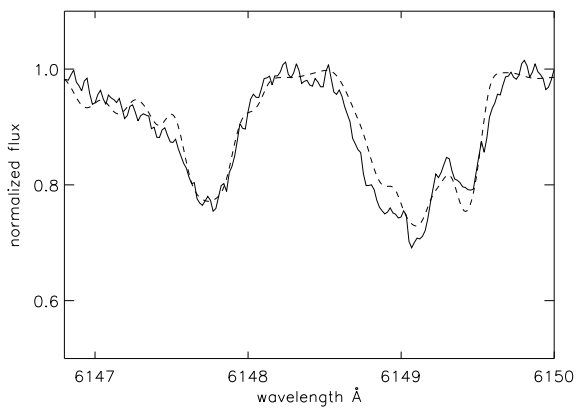}
\caption{\label{sy57040} Observed (solid) and synthetic (dashed) spectral region 
for HD\,57040. The synthetic spectrum was calculated for a magnetic field of 
7.8\,kG. The two strongest lines belong to Fe\,\textsc{ii} 6147.741\,\AA\ and 
6149.258\,\AA. The partially split line at 6149.258\,\AA\ is blended with weaker 
lines of Sm\,\textsc{ii} and Ce\,\textsc{ii}.}
\end{center}
\end{figure}

This star has a peculiar spectrum with a strong magnetic field. Our FEROS spectrum 
shows only a hint of magnetic splitting for the Fe\,\textsc{ii} 6149.258\,\AA\ 
line, whereas another spectrum obtained with UVES and the VLT shows Zeeman 
splitting. The spectrum has strong lines of Nd\,\textsc{ii}, Nd\,\textsc{iii}, 
Eu\,\textsc{ii} and Ce\,\textsc{ii}. With $T_{\rm eff} =7500$\,K the star is a 
promising candidate to be an roAp star. In Fig.\,\ref{sy57040} part of a UVES 
spectrum of HD\,57040 is shown with magnetic splitting in the Fe\,\textsc{ii} 
6149.258\,\AA\ line. A nearby spectral range with an example of Zeeman patterns is 
shown in Fig.\,1 of \citet{Mathys97}.

\begin{figure}
\begin{center}
\hfil \epsfxsize 8.2cm\epsfbox{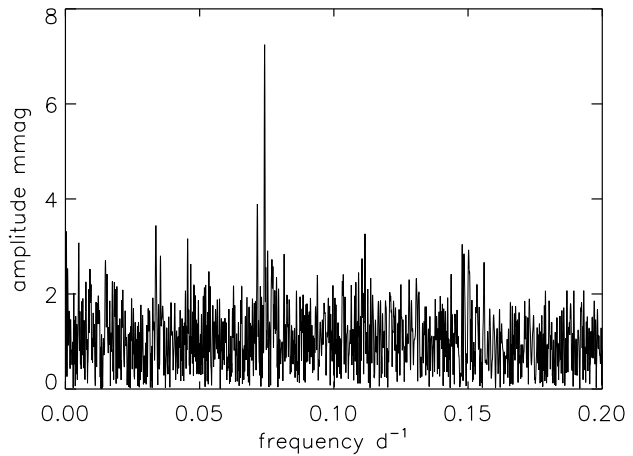}
\caption{\label{ft57040} An amplitude spectrum of ASAS photometry for HD\,57040 
with a peak corresponding to the rotational period of 13.474\,d. }
\end{center}
\end{figure}

Spectral lines in HD\,57040 also show rotational broadening. ASAS photometry 
reveals for this star a probable rotational period of 13.474\,d. 
Fig.\,\ref{ft57040} illustrates the amplitude spectrum of the ASAS photometry 
with a significant peak corresponding to this period.

While two photometric attempts to detect pulsation in this star by \citet{Mart94} 
were unsuccessful, pulsations with small amplitude may still exist. For example, 
low pulsation amplitude may be found using satellite photometry (e.g. 
\citealt{Kurtz11}). We obtained 34 UVES spectra to test the star for rapid radial 
velocity variations. The analysis of these spectra will be presented in a separate 
paper.

\subsubsection{HD\,61513}

\begin{figure}
\begin{center}
\hfil \epsfxsize 8.2cm\epsfbox{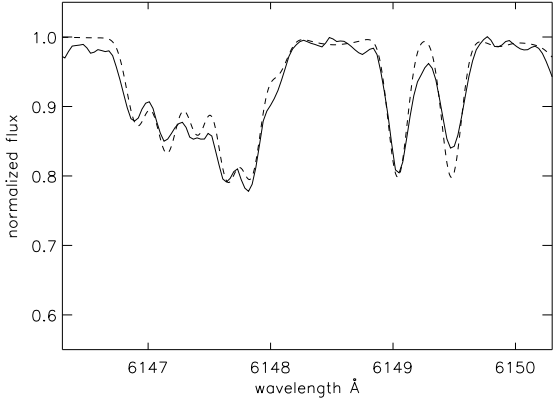}
\caption{\label{sy61513} Observed (solid) and synthetic (dashed) spectra for 
HD\,61513. The synthetic spectrum was calculated for a magnetic field strength of 
9.2\,kG. The profile of the Fe\,\textsc{ii} 6149.258\,\AA\ line shows doublet 
splitting.}
\end{center}
\end{figure}

This star is among the faintest ($V = 10.15$) and hottest ($T_{\rm eff} = 
10\,000$\,K) stars we observed with FEROS. Lines of Nd\,\textsc{iii}, 
Eu\,\textsc{ii}, Ce\,\textsc{ii} and some other rare earth elements are present in 
the spectrum at moderate strength for an Ap star. Some rotational broadening is 
present corresponding to $v \sin i = 7.0 \pm 1.5$\,km\,s$^{-1}$. The magnetic 
field is strong and many lines show Zeeman splitting. Components of the 
Fe\,\textsc{ii} 6149.258\,\AA\ line are clearly resolved as seen in 
Fig.\,\ref{sy61513}. Direct measurements of the magnetic field from this line and 
by fitting with synthetic spectrum calculated with {\small SYNTHMAG} give similar 
results, 9.2\,kG.

\subsubsection{HD\,70702}

This is another hot star ($T_{\rm eff} = 9800$\,K) in our target list. The 
spectrum is peculiar with rare earth element lines present, including 
Nd\,\textsc{iii} and Eu\,\textsc{ii}, but many lines are rather shallow.

The star shows a very strong magnetic field of 15\,kG, which was not easy to 
recognize because of significant rotational broadening, $v \sin i = 17.0 \pm 
1.5$\,km\,s$^{-1}$. The rotational period should be no longer than several days. A 
comparison of the observed and synthetic profiles allowed us to distinguish 
blending and Zeeman splitting in the Fe\,\textsc{ii} 6149.258\,\AA\ line and 
determine a significant magnetic field in this star.

\begin{figure}
\begin{center}
\hfil \epsfxsize 8.2cm\epsfbox{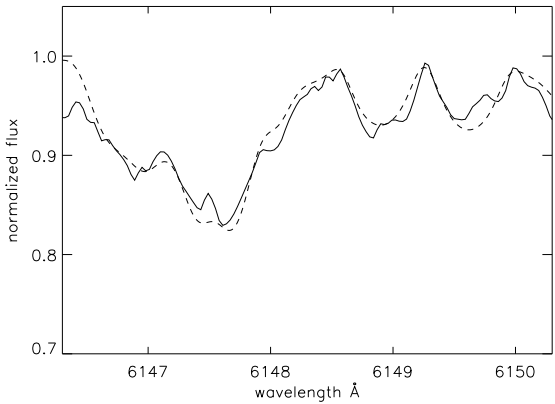}
\caption{\label{sy70702} Observed (solid) and synthetic (dashed) spectra for 
HD\,70702. The profile of the Fe\,\textsc{ii} 6149.258\,\AA\ line shows huge 
Zeeman splitting. The synthetic spectrum was calculated for a magnetic field of 
15\,kG.}
\end{center}
\end{figure}

Zeeman splitting in the Fe\,\textsc{ii} 6149.258\,\AA\ line is presented in 
Fig.\,\ref{sy70702}. The split components show a complex doublet structure. This 
may be explained by high noise level or blending. A nonuniform distribution of 
iron in the line formation region combined with different field strengths also may 
be responsible for the asymmetry of the Zeeman patterns.

Some other spectral lines also show Zeeman structure, as is confirmed by synthetic 
calculations for magnetic field strengths in the range $14 - 16$\,kG. Zeeman 
splitting is visible, for example, in Cr\,\textsc{ii} 5046.940\,\AA, in 
Fe\,\textsc{ii} 6238.392\,\AA\ and in Eu\,\textsc{ii} 6437.640\,\AA. Most other 
lines demonstrate just magnetic broadening.

With such a strong magnetic field this star is an important target for further 
observations and magnetic field analysis.

\subsubsection{HD\,76460}

\begin{figure}
\begin{center}
\hfil \epsfxsize 8.0cm\epsfbox{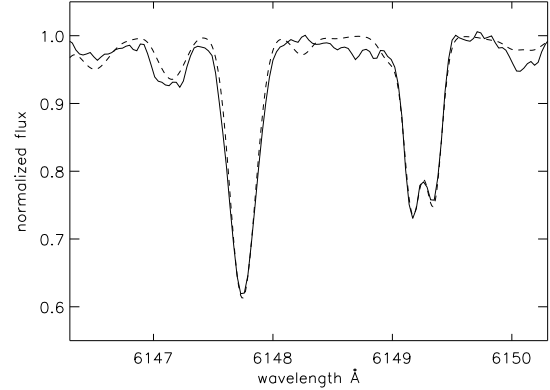}
\caption{\label{sy76460} Observed (solid) and synthetic (dashed) spectra for 
HD\,76460. The profile of the Fe\,\textsc{ii} 6149.258\,\AA\ line shows partial 
Zeeman splitting. The synthetic spectrum was calculated for a magnetic field of 
3.7\,kG.}
\end{center}
\end{figure}

This star has a peculiar spectrum with narrow lines. Those of Ba\,\textsc{ii} are 
very strong while rare earth element lines, including Nd\,\textsc{iii} and 
Eu\,\textsc{ii}, have moderate intensities. The doublet line of Li\,\textsc{ii} 
6708\,\AA\ is also present in the spectrum. The magnetic field is strong enough 
for partial splitting of the Fe\,\textsc{ii} 6149.258\,\AA\ line as can be seen in 
Fig.\,\ref{sy76460}. The line of Fe\,\textsc{i} 6336.823\,\AA\ also shows partial 
doublet splitting. We measure the field strength to be 3.7\,kG.

\subsubsection{HD\,81588}

\begin{figure}
\begin{center}
\hfil \epsfxsize 8.2cm\epsfbox{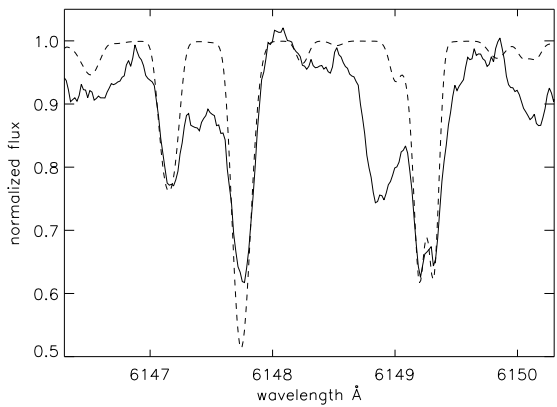}
\caption{\label{sy81588} Observed (solid) and synthetic (dashed) spectra for 
HD\,81588. The profile of the Fe\,\textsc{ii} 6149.258\,\AA\ line shows partial 
Zeeman splitting. The synthetic spectrum was calculated for a magnetic field of 
2.4\,kG.}
\end{center}
\end{figure}

The star was first observed with FEROS and showed a highly peculiar spectrum with 
strong rare earth element lines of Nd\,\textsc{iii} and Pr\,\textsc{iii}. The 
doublet Li\,\textsc{ii} 6708\,\AA\ line was also detected, but it is a blend with 
lines of Ce\,\textsc{ii} and Sm\,\textsc{ii}. The spectral lines are sharp and 
narrow, but the FEROS resolution was not sufficient to see magnetic splitting in 
the Fe\,\textsc{ii} 6149.258\,\AA\ line. A UVES spectrum showed small partial 
splitting as seen in Fig.\,\ref{sy81588}. The line of Fe\,\textsc{i} 
6336.823\,\AA\ also demonstrates partial doublet splitting for the UVES spectrum. 
In the FEROS spectrum this line has only magnetic broadening.

The fundamental parameters of HD\,81588 determined from photometry and 
spectroscopy are similar to some known roAp stars. \citet{Mart94} observed this 
star five times photometrically searching of rapid oscillations and did not detect 
any.

\subsubsection{HD\,88241}

\begin{figure}
\begin{center}
\hfil \epsfxsize 8.2cm\epsfbox{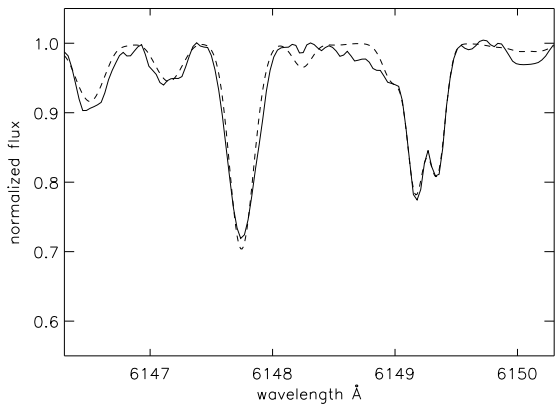}
\caption{\label{sy88241} Observed (solid) and synthetic (dashed) spectra for 
HD\,88241. The profile of the Fe\,\textsc{ii} 6149.258\,\AA\ line shows partial 
Zeeman splitting. The synthetic spectrum was calculated for a magnetic field of 
3.6\,kG.}
\end{center}
\end{figure}

The spectrum of this magnetic star has very strong lines of Ba\,\textsc{ii} and 
good lines of Nd\,\textsc{iii}. Other rare earth element lines such as 
Eu\,\textsc{ii} and Gd\,\textsc{ii} are also present. This star has a strong 
Li\,\textsc{i} 6708\,\AA\ doublet. The Fe\,\textsc{ii} 6149.258\,\AA\ line shows 
partial Zeeman splitting as seen in Fig.\,\ref{sy88241}. The line of 
Fe\,\textsc{i} 6336.823\,\AA\ also demonstrates partial doublet splitting. The 
synthetic calculations and fitting yield a magnetic field 3.6\,kG. Other lines
with large Land\'e factors also show magnetic broadening.

\subsubsection{HD\,158450}

Two spectra of this peculiar and strongly magnetic star were obtained with FEROS. 
The spectral lines show magnetic Zeeman splitting or broadening and rotational 
broadening. Rare earth element lines found in the spectra include Nd\,\textsc{iii} 
and Eu\,\textsc{ii}, among others. These lines are relatively weak in comparison 
with most peculiar stars.

\begin{figure}
\begin{center}
\hfil \epsfxsize 8.2cm\epsfbox{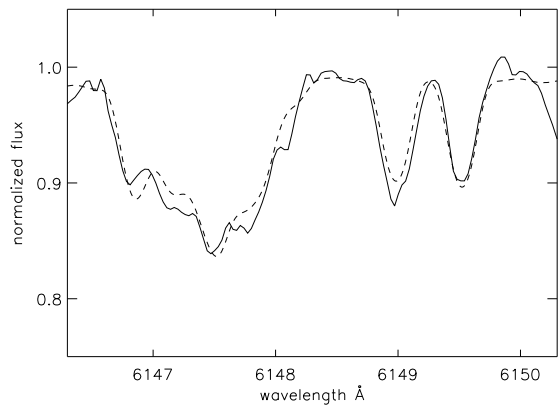}
\caption{\label{sy158450} Observed (solid) and synthetic (dashed) spectra of the 
second FEROS spectrum obtained for HD\,158450. The profile of the Fe\,\textsc{ii} 
6149.258\,\AA\ line shows Zeeman splitting. The synthetic spectrum was calculated 
for a magnetic field of 11.2\,kG.}
\end{center}
\end{figure}

\begin{figure}
\begin{center}
\hfil \epsfxsize 8.2cm\epsfbox{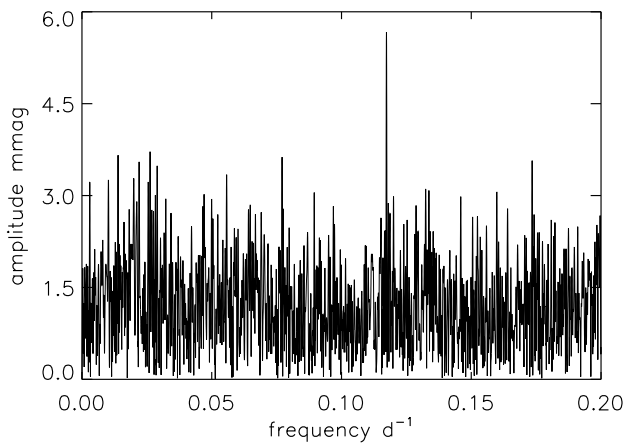}
\caption{\label{ft158450} An amplitude spectrum of the ASAS photometry for 
HD\,158450 with a peak corresponding to a rotational period of 8.524\,d. } 
\end{center}
\end{figure}

\begin{figure}
\begin{center}
\hfil \epsfxsize 8.2cm\epsfbox{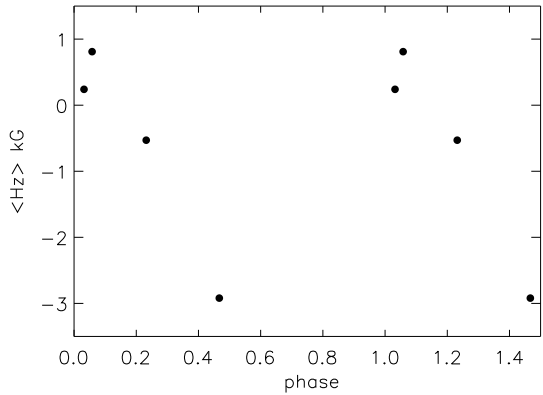}
\caption{\label{be158450} The longitudinal magnetic field of HD\,158450  from 
\citep{Kudr06} with phase of 8.524\,-d period. }
\end{center}
\end{figure}

Direct measurements of the field from split Zeeman components of the 
Fe\,\textsc{ii} 6149.258\,\AA\ line reveal a magnetic field modulus of 11.9\,kG 
for one spectrum and 11.2\,kG for a second observation. In Fig.\,\ref{sy158450} 
the spectral region with the Fe\,\textsc{ii} 6149.258\,\AA\ line is shown together 
with a synthetic spectrum calculated for an 11.2\,kG field. Despite the strong 
field only a small number of lines demonstrate Zeeman splitting. The main reason 
is rotational broadening. The Fe\,\textsc{ii} 6238.392\,\AA\ line has doublet 
splitting corresponding to a magnetic field of 9.9\,kG. The longitudinal magnetic 
field was found to vary between $-2.92$ to $+0.81$\,kG over several days 
\citep{Kudr06}. Only four observations of this star have been published, and they 
are consistent with the rotational period found from the photometry.
 Fig.\,\ref{ft158450} presents an amplitude spectrum of the ASAS photometry, 
while Fig.\,\ref{be158450} shows the longitudinal 
magnetic field phased with the photometric 8.524-d period.

 In the Catalogue of Components of 
Double and Multiple stars \citep{Dommanget02}, this star is noted to be a binary 
star with a faint 10.2\,mag component at distance of 0.6\,arcsec.

\citet{Mart94} observed HD\,158450 photometrically for about 2\,h to search for 
rapid variations. This result was uncertain and further similar observations would 
be useful.

\subsubsection{HD\,162316}

\begin{figure}
\begin{center}
\hfil \epsfxsize 8.2cm\epsfbox{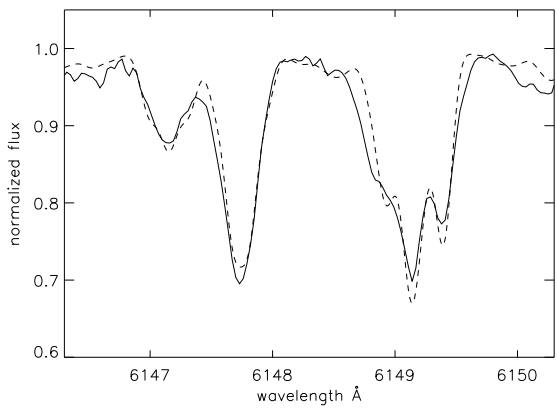}
\caption{\label{sy162316} Observed (solid) and synthetic (dashed) spectra for 
HD\,162316. the synthetic spectrum was calculated for a magnetic field of 6.0\,kG. 
It can be seen that the Fe\,\textsc{ii} 6149.258\,\AA\ line shows partial Zeeman 
splitting. This line is a blend with a strong line of Sm\,\textsc{ii} 
6149.060\,\AA.}
\end{center}
\end{figure}

\begin{figure}
\begin{center}
\hfil \epsfxsize 8.2cm\epsfbox{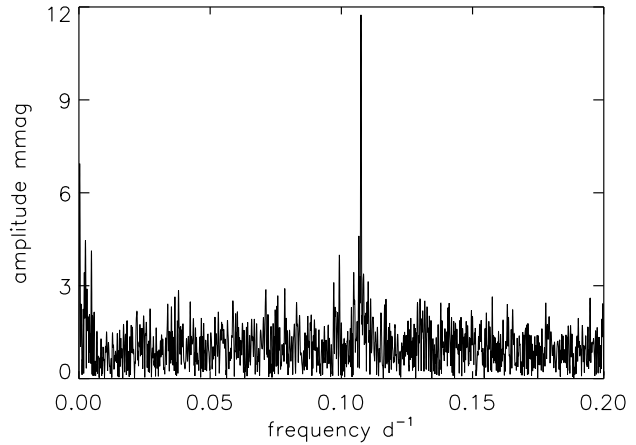}
\caption{\label{ft162316} Fourier transform of ASAS photometry for HD\,162316 with 
a peak corresponding to a rotational period of 9.304\,d. }
\end{center}
\end{figure}

This star has very strong lines of Nd\,\textsc{iii} while other lines of rare 
earth elements such as Eu\,\textsc{ii} are also present. Zeeman splitting is 
visible in the Fe\,\textsc{ii} 6149.258\,\AA\ line as shown in 
Fig.\,\ref{sy162316}. This line is a blend, but comparison with a synthetic 
spectrum proves the presence of a magnetic field. The Fe\,\textsc{i} 
6336.823\,\AA\ line also shows a doublet structure of Zeeman components. Lines 
with low Land\'e factors are narrower than other lines. ASAS photometry shows 
variability with a probable rotation period of 9.304\,d. A clear peak is visible 
in the amplitude spectrum shown in Fig.\,\ref{ft162316}.

\subsubsection{HD\,168767}

\begin{figure}
\begin{center}
\hfil \epsfxsize 8.2cm\epsfbox{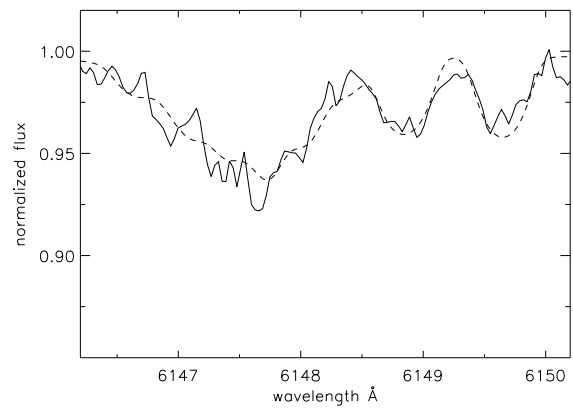}
\caption{\label{sy168767} Observed (solid) and synthetic (dashed) spectra for 
HD\,168767. Zeeman splitting is visible for the Fe\,\textsc{ii} 6149.258\,\AA\ 
line when compared with a synthetic spectrum calculated for a magnetic field of 
16.5\,kG. }
\end{center}
\end{figure}

This star shows a peculiar spectrum, but most of the metal lines are weak and 
shallow with rotational and magnetic broadening. The spectrum has rather weak 
lines of Nd\,\textsc{iii}, Pr\,\textsc{iii} and Eu\,\textsc{ii}. The magnetic 
field can be recognised from splitting of the Fe\,\textsc{ii} 6149.258\,\AA\ line, 
although it was not obvious and required a synthetic spectrum for comparison as 
shown in Fig.\,\ref{sy168767}.

The Fe\,\textsc{ii} 6238.392\,\AA\ line also shows doublet Zeeman splitting, which 
fits well when compared to a synthetic spectrum calculated for a magnetic field of 
16.5\,kG. Despite this strong field, most other lines do not show splitting 
because of relatively rapid rotation with $v \sin i = 14.0 \pm 1.5$\,km\,s$^{-1}$. 
With a strong magnetic field and relatively short rotational period (not more than 
several
 days given the relatively high $v \sin i$), this star is one of the most 
interesting targets
 for  future observations. We have only one spectrum; observations at other
 rotational phases may reveal an even stronger magnetic field.

\subsubsection{HD\,177268}

\begin{figure}
\begin{center}
\hfil \epsfxsize 8.2cm\epsfbox{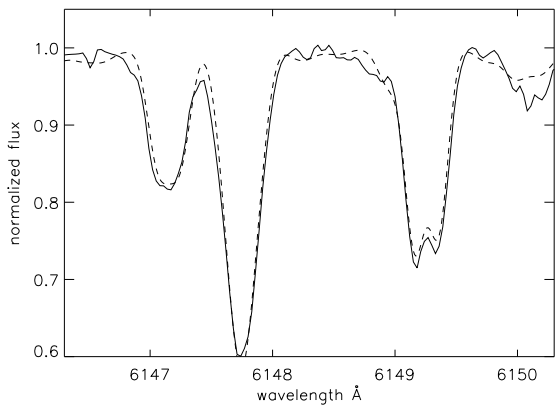}
\caption{\label{sy177268} Observed (solid) and synthetic (dashed) spectra for 
HD\,177268. The partial Zeeman splitting is visible for the Fe\,\textsc{ii} 
6149.258\,\AA\ line. The synthetic spectrum was calculated for a magnetic field of 
4.1\,kG }
\end{center}
\end{figure}

This star was observed with FEROS three times as it was not clear whether it shows 
magnetic splitting. Moderate intensity spectral lines of rare earth elements are 
present in the spectrum. The partial Zeeman splitting of the Fe\,\textsc{ii} 
6149.258\,\AA\ line is visible for two spectra and shows some hint of splitting 
for the third. This means that the magnetic field is variable with an unknown 
rotation period. ASAS photometry did not give a clue to a possible period. 
Fig.\,\ref{sy177268} demonstrates a portion of one of the spectra together with a 
synthetic spectrum. Another Fe\,\textsc{i} 6336.823\,\AA\ line also shows partial 
Zeeman splitting, thus supporting the discovery of a magnetic field in this star.
The star has physical parameters similar to known roAp stars, which led 
\citet{Mart94} 
to search photometrically for rapid oscillations,  but no 
evidence for pulsation was found.

\subsubsection{HD\,179902}

\begin{figure}
\begin{center}
\hfil \epsfxsize 8.2cm\epsfbox{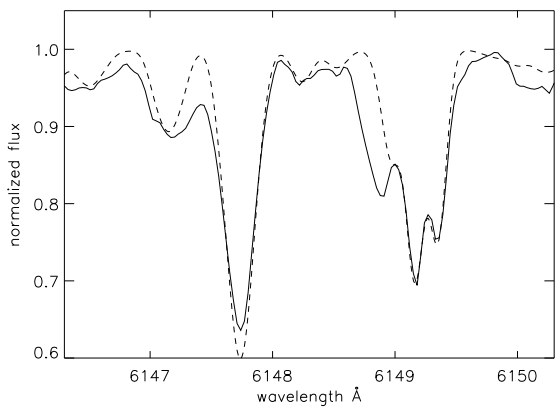}
\caption{\label{sy179902} Observed (solid) and synthetic (dashed) spectra for 
HD\,179902. The synthetic spectrum was calculated for a magnetic field 3.9\,kG. 
Partial Zeeman splitting is visible for the Fe\,\textsc{ii} 6149.258\,\AA\ line. }
\end{center}
\end{figure}

This is another star with very strong lines of Nd\,\textsc{iii}. Other strong rare 
earth element lines are also found in the spectrum, including Pr\,\textsc{iii} and 
Eu\,\textsc{ii}. The spectrum is reminiscent of HD\,162316 with similar 
peculiarities. Even the lines of Fe\,\textsc{ii} 6149.258\,\AA, which are heavily 
blended in both stars, show similar blending and splitting. The splitting in this 
line is partial, so it requires a {\small SYNTHMAG} synthetic calculation to 
estimate the magnetic field strength. Fig.\,\ref{sy179902} shows the profile of 
this iron line in comparison with a synthetic spectrum. The magnetic field also is 
confirmed by partial doublet splitting of the Fe\,\textsc{i} 6336.823\,\AA\ line. 
This star was observed twice with FEROS with a 6-d gap, but no significant 
spectral variability was detected. The magnetic field also did not change within 
the errors for these two spectra. In contrast with HD\,162316, we did not find any 
significant variability for HD\,179902 using ASAS photometry.

\subsubsection{HD\,184120}

\begin{figure}
\begin{center}
\hfil \epsfxsize 8.2cm\epsfbox{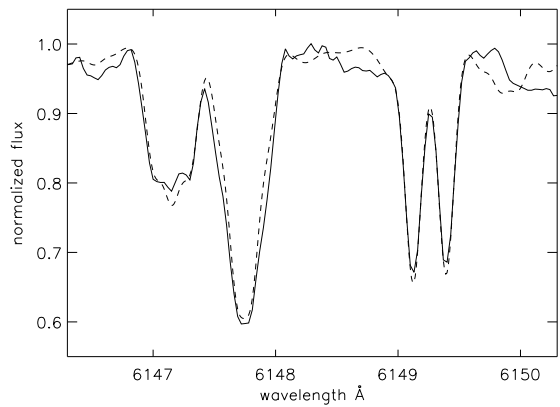}
\caption{\label{sy184120} Observed (solid) and synthetic (dashed) spectra for 
HD\,184120. The synthetic spectrum was calculated for a magnetic field 5.7\,kG. 
Zeeman splitting is clearly visible for the Fe\,\textsc{ii} 6149.258\,\AA\ line. }
\end{center}
\end{figure}

This peculiar star shows rather moderate or weak lines of Nd\,\textsc{iii}, 
Eu\,\textsc{ii} and some other rare earth elements in the spectrum. The magnetic 
field is strong and Zeeman components of the Fe\,\textsc{ii} 6149.258\,\AA\ line 
are split, as can be seen in Fig.\,\ref{sy184120}. Some other lines with large 
Land\'e factors also show Zeeman splitting, mostly partial. Many lines demonstrate 
magnetic broadening. This star was observed with FEROS twice and both spectra are 
similar with no significant spectral variability.
\citet{Mart94} did not find any rapid photometric variability, even though
 the spectrum and physical parameters of this star are similar to roAp stars.

\subsubsection{HD\,185204}

\begin{figure}
\begin{center}
\hfil \epsfxsize 8.2cm\epsfbox{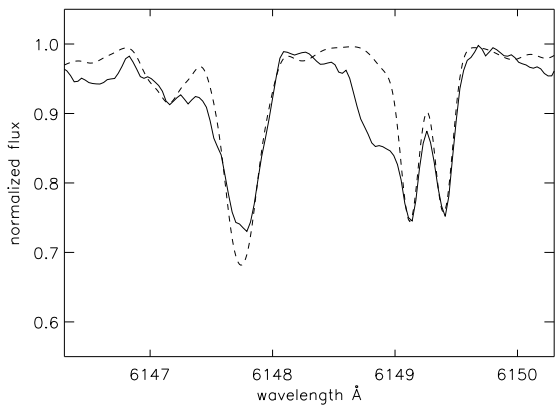}
\caption{\label{sy185204} Observed (solid) and synthetic (dashed) spectra for 
HD\,185204. The synthetic spectrum was calculated for a magnetic field of 5.9\,kG. 
Zeeman splitting is clearly visible for the Fe\,\textsc{ii} 6149.258\,\AA\ line, 
which is a blend with an unknown line. }
\end{center}
\end{figure}

This magnetic star has a peculiar spectrum with very strong lines of 
Nd\,\textsc{iii} and good lines of Pr\,\textsc{iii}, Eu\,\textsc{ii} and some 
other rare earth elements. This peculiar star was observed twice; both spectra are 
similar. Zeeman splitting is visible in the Fe\,\textsc{ii} 6149.258\,\AA\ line, 
as can be seen in Fig.\,\ref{sy185204}. The line of Fe\,\textsc{i} 6336.823\,\AA\ 
also shows partial doublet splitting and partial splitting is also visible for 
many other lines across the spectrum.

This star is a promising target for searching for rapid oscillations. 
\citet{Mart94} observed it twice photometrically, but pulsations were not found. 
Additional precise high time resolution observations will be useful to test 
further if the star pulsates with low amplitude.

\subsubsection{HD\,191695}

\begin{figure}
\begin{center}
\hfil \epsfxsize 8.2cm\epsfbox{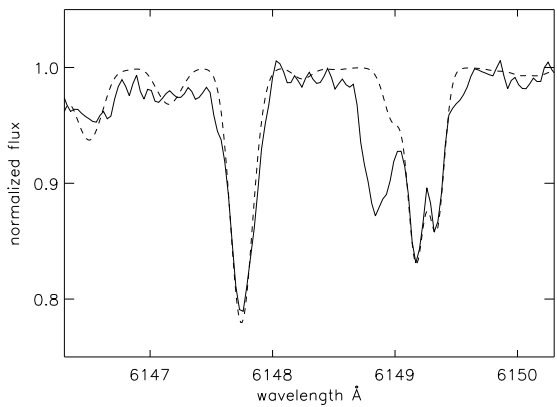}
\caption{\label{sy191695} Observed (solid) and synthetic (dashed) spectra for 
HD\,191695. The synthetic spectrum was calculated for a magnetic field of 3.4\,kG. 
Zeeman splitting is clearly visible for the Fe\,\textsc{ii} 6149.258\,\AA\ line, 
which is a blend with Sm\,\textsc{ii} 6149.060\,\AA\ and with another unknown 
line. }
\end{center}
\end{figure}

This peculiar star shows very strong lines of Nd\,\textsc{iii}, Pr\,\textsc{iii} 
and Ba\,\textsc{ii}, but lines of Eu\,\textsc{ii} are rather weak. The magnetic 
field is not very strong and only partial Zeeman splitting is present in the 
Fe\,\textsc{ii} 6149.258\,\AA\ line as can be seen in Fig.\,\ref{sy191695}. The 
presence of a magnetic field is also supported by partial doublet splitting of the 
Fe\,\textsc{i} 6336.823\,\AA\ line. \citet{Nelson_Kreidl93} did not find rapid 
oscillations in HD\,191695 above a noise level of 1\,mmag, but this cool Ap star 
is still a good target to search for low amplitude pulsation with high precision, 
using both spectroscopic and photometric high time resolution observations.

\section{Discussion and conclusions}

\begin{table*}
\begin{center}
\caption{List of detected magnetic stars. The columns give the star�s name, 
magnitude, the Modified Julian Date (MJD) of the start of each exposure, exposure 
time, and the magnetic field modulus, effective temperature and projected 
rotational velocity. The error estimates for the determined parameters are 
described in the text.}
\begin{tabular}{rrrccrc}
\hline
\multicolumn{1}{c}{Star}  &
\multicolumn{1}{c}{$V$}  &
\multicolumn{1}{c}{MJD}  &
\multicolumn{1}{c}{exposure}  &
\multicolumn{1}{c}{magnetic field}  &
\multicolumn{1}{c}{$T_{\rm eff}$}   &
\multicolumn{1}{c}{$v \sin i$}   \\
\multicolumn{1}{c}{}    &
\multicolumn{1}{c}{}    &
\multicolumn{1}{c}{}    &
\multicolumn{1}{c}{time (s)}    &
\multicolumn{1}{c}{modulus (kG)}    &
\multicolumn{1}{c}{}   &
\multicolumn{1}{c}{ km\,s$^{-1}$ }   \\
\hline
HD\,3988   &  8.4 &    54686.37371 &    321 &    2.7 $\pm$  0.2   &     7200   &  
3.0  \\
           &      &    54687.37890 &    540 &    2.5 $\pm$  0.2   &            &       
\\
           &      &    54690.29598 &    600 &    2.7 $\pm$  0.2   &            &       
\\
           &      &    54691.29863 &    700 &    2.7 $\pm$  0.2   &            &       
\\
HD\,57040  &  9.2 &    54444.28124 &   3480 &    7.5 $\pm$  0.4   &     7600   &  
5.5  \\
HD\,61513  & 10.1 &    54870.15212 &   1200 &    9.2 $\pm$  0.1   &    10000   &  
7.0  \\
HD\,70702  & 8.5  &    54141.10660 &    420 &   15.0 $\pm$  0.6   &     9800   &  
17.0 \\
HD\,76460  & 9.8  &    55227.20701 &   1100 &    3.6 $\pm$  0.2   &     7200   &  
3.0  \\
HD\,81588  & 8.5  &    54515.21196 &   3340 &    2.4 $\pm$  0.2   &     7400  &  
3.0    \\
HD\,88241  & 8.6  &    55228.15571 &    500 &    3.6 $\pm$  0.2   &     7000  &  
3.5   \\
HD\,158450 & 8.5  &    54686.01386 &    371 &   11.9 $\pm$  0.3  &     8000  &  
7.5   \\
           &      &    55022.27403 &    900 &   11.2 $\pm$  0.3   &           &        
\\
HD\,162316 & 9.4  &    55029.17408 &   1100 &    6.0 $\pm$  0.2   &     7600  &  
3.0   \\
HD\,168767 & 8.7  &    54686.11474 &    480 &   16.5 $\pm$  0.6   &     7600  & 
14.0   \\
HD\,177268 & 9.1  &    54689.24174 &    950 &    3.7 $\pm$  0.2   &     7800  &  
3.5   \\
           &      &    55023.27249 &   1100 &    3.9 $\pm$  0.2   &           &        
\\
           &      &    55029.18944 &   1100 &    4.0 $\pm$  0.2   &           &        
\\
HD\,179902 & 10.0 &    55023.30541 &   1200 &    3.7 $\pm$  0.2   &     7200  &  
3.0   \\
           &      &    55029.26619 &   2400 &    3.9 $\pm$  0.2  &           &        
\\
HD\,184120 & 10.2 &    55023.33752 &   1200 &    5.8 $\pm$  0.1   &     7400  &  
4.0   \\
           &      &    55030.31810 &   2400 &    5.7 $\pm$  0.1   &           &         
\\
HD\,185204 & 9.6  &    54690.24203 &   1200 &    5.7 $\pm$  0.2   &     7400  &  
4.0     \\
           &      &    54688.25639 &   1023 &    5.7 $\pm$  0.2   &           &         
\\
HD\,191695 & 9.9  &    55023.36940 &   1200 &    3.4 $\pm$  0.2   &     7000  &  
2.7    \\
           &      &    55028.36782 &   1200 &    3.0 $\pm$  0.2   &           &         
\\
\hline
\end{tabular}
\label{mag_f2}
\end{center}
\end{table*}

Measured magnetic fields in Ap stars show significant variability with rotation 
period. The magnetic oblique rotator model explains this variability as an aspect 
effect of the observed star. Most stars presented here were observed in just one 
or two rotation phases and require more observations over their rotation periods 
to establish how strong their magnetic fields are and to determine their 
geometries. It is especially interesting to observe stars with strong magnetic 
fields like those of HD\,70702 and HD\,168767. The extreme values of the fields in 
these and other stars we have presented here may be higher in the other rotational 
phases.

Table\,1 shows the results of magnetic field measurements together with other 
determined parameters of the stars. The standard deviation for magnetic field 
measurements for well resolved Zeeman components with high signal to noise ratios 
in the spectra is about $100$\,G, while for partially split lines, blended lines 
and for spectra with high noise it is in the range $200 - 500$\,G. The effective 
temperatures in this table are based mostly on fitting the observed H$\alpha$ 
profiles with synthetic profiles. At best we estimate the error on $T_{\rm eff}$ 
to be $200 - 300$\,K. For most of the stars studied the difference between 
photometric and spectroscopic effective temperatures is less than 500\,K, although 
for a few cases, especially for hotter stars, this difference is larger.

Our $v \sin i$ parameter determination has a precision of about $1 - 
1.5$\,km\,s$^{-1}$. While for some stars a $v \sin i$ value of 3\,km\,s$^{-1}$ was 
obtained, this is just the lower limit for the FEROS resolution. Considering that 
the magnetic field stabilises the stellar atmosphere, we used a value of zero for 
the microturbulence and macroturbulence velocities in all calculations. With ASAS 
photometry \citep{Pojmanski02} and using the Period04 program by 
\citet{lenzetal05} we tested stars from Table\,1 and found rotation periods for 
three of them.

\begin{figure}
\begin{center}
\hfil \epsfxsize 8.2cm\epsfbox{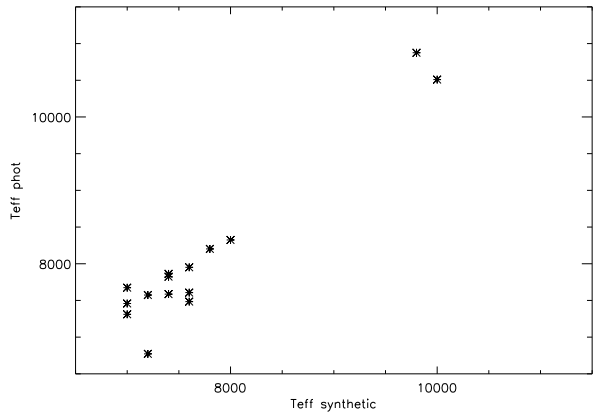}
\caption{\label{sy_ph_teff} Correlation of effective temperatures determined from 
photometry and spectroscopy.}
\end{center}
\end{figure}

A correlation of the effective temperatures obtained by photometry and 
spectroscopy is shown in Fig.\,\ref{sy_ph_teff}. The agreement between effective 
temperatures obtained with different methods is mostly acceptable, while in a few 
cases further analysis is needed to resolve the discrepancy. We prefer to use the 
effective temperature obtained with spectroscopic analysis, since the photometric 
calibrations are known to be problematic for extremely peculiar stars, as a 
consequence of line blocking.

One of the fundamental questions of the physics Ap stars concerns the relation 
between magnetic field strength and rotational period. The Ap stars rotate much 
more slowly than normal stars with the same effective temperature. Typically the 
rotation periods of magnetic Ap stars range from several days to many years, and 
even decades. The magnetic field is responsible for braking Ap stars. 

\begin{figure}
\begin{center}
\hfil \epsfxsize 8.2cm\epsfbox{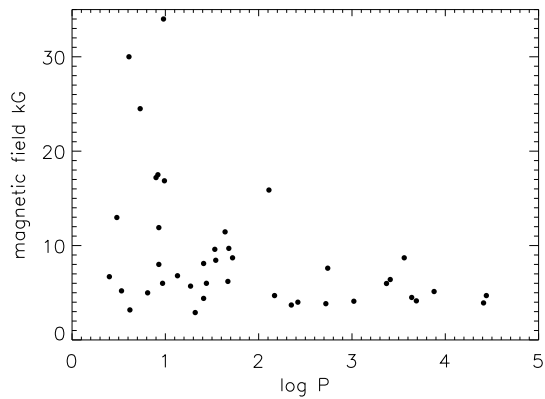}
\caption{\label{bs_p} Logarithm of rotational period versus extrema of magnetic 
field modulus for magnetic stars.}
\end{center}
\end{figure}

To examine this relationship further here, we 
collected more magnetic field and period values for Ap stars from the 
literature. A graph for rotational period as a function of extrema of magnetic 
field modulus for a sample of magnetic stars is presented in Fig.\,\ref{bs_p}. 
Data for 30 stars in this figure were taken from \citet{Mathys97}. For other stars 
the data were obtained from \citet{Elkin10b}, \citet{Freyhammer08}, 
\citet{Hubrig09}, \citet{Mathys07}, \citet{Ryab06} and the current paper. This 
figure demonstrates that the stars with strongest magnetic fields typically have 
rotational periods between 5 and 10\,d. This is also supported by our observations 
of two stars HD\,70702 and HD\,168767 presented in this paper. Both stars have a 
very strong field and relatively high projected rotational velocities,  which 
suggests that rotational period should not be more than several days. The lack of 
stars with very strong fields and with periods less than 5\,d may be at least 
partly explained by selection effects, as the fast rotators have wider spectral 
lines and even for fields more than 10\,kG the Zeeman components are not resolved. 
Spectropolarimetric techniques would be useful for searching for longitudinal 
fields among the fast rotators. A good example is the star NGC\,2244-334 
\citep{Bagnulo04}, which shows a very strong longitudinal field and has wide 
spectral lines with magnetic and rotational broadening. This star also should have 
a relatively short rotational period.

\citet{Mathys97} suggested a possible anticorrelation between the mean magnetic 
field modulus and stellar rotation period. Figure\,50 from \citet{Mathys97} 
differs from our Fig.\,\ref{bs_p} because we have included several stars with very 
strong fields that were found subsequent to \citet{Mathys97}. The root-mean-square 
longitudinal magnetic fields as a function of rotation period were studied by 
\citet{hubrig07} on a wider sample of Ap stars. They showed that the stars with 
strongest longitudinal field generally show periods less than 10\,d.

Together with previous papers (\citealt{Freyhammer08}; \citealt{Elkin10a}; 
\citealt{Elkin11}) we have found a total of 34 new magnetic stars with resolved 
and partly resolved Zeeman components using high resolution spectra from our FEROS 
survey of cool Ap stars. Among them we found several stars with a mean magnetic 
field modulus more than 10\,kG. Considering the number of observed Ap stars in 
this survey we can estimate that the proportion of stars with resolved Zeeman 
components is slightly less than $10$\,percent. \citet{Mathys04} noted only 47 
magnetic Ap/Bp stars with resolved Zeeman splitting. Our discovery of 34 stars 
with clear 
Zeeman splitting found among cool Ap stars is a significant contribution for  
further study of this type of star. This comes from our survey of more than three 
hundred cool 
Ap stars from which we also found several stars with very strong magnetic fields. 
The strongest 
field of 30\,kG was found in HD\,75049. Most probably this value is close to a 
physical limit for the observed magnetic field in cool Ap stars.

\section{Acknowledgements}
DWK and VGE acknowledge support for this work from the Science and Technology 
Facilities Council (STFC). This research has made use of SIMBAD database, operated 
at CDS, Strasbourg, France.

{}

\end{document}